# Evolution of the strange-metal scattering in momentum space of electron-doped $La_{2-x}Ce_xCuO_4$

Cenyao Tang[1,2,3,*], Zefeng Lin[1,2,*], Shunye Gao[1,2], Jin Zhao[1,2], Xingchen Guo[1,2], Zhicheng Rao[1,2], Yigui Zhong[1,2], Xilin Feng[1,2], Jianyu Guan[1,2], Yaobo Huang[4], Tian Qian[1,5], Kun Jiang[1,2] ✉, Kui Jin[1,2,5] ✉, Yujie Sun[6,1] ✉ and Hong Ding[1,2,7]

[1] Beijing National Laboratory for Condensed Matter Physics and Institute of Physics, Chinese Academy of Sciences, Beijing 100190, China
[2] University of Chinese Academy of Sciences, Beijing 100049, China
[3] State Key Laboratory of Surface Physics and Department of Physics, Fudan University, Shanghai 200433, China.
[4] Shanghai Advanced Research Institute, Chinese Academy of Sciences, Shanghai 201204, China
[5] Songshan Lake Materials Laboratory, Dongguan, Guangdong 523808, China
[6] Department of Physics, Southern University of Science and Technology, Shenzhen 518055, China
[7] Tsung-Dao Lee Institute & School of Physics and Astronomy, Shanghai Jiao Tong University, Shanghai 200240, China

*These authors contributed equally to this work.
✉Corresponding authors: jiangkun@iphy.ac.cn; kuijin@iphy.ac.cn; sunyj@sustech.edu.cn;

**Abstract**
The linear-in-temperature resistivity is one of the important mysteries in the strange metal state of high-temperature cuprate superconductors. To uncover this anomalous property, the energy-momentum-dependent imaginary part of the self-energy Im $\Sigma(k, \omega)$ holds the key information. Here we perform systematic doping, momentum, and temperature-dependent angle-resolved photoemission spectroscopy measurements of electron-doped cuprate $La_{2-x}Ce_xCuO_4$ and extract the evolution of the strange metal scattering in momentum space. At low doping levels and low temperatures, Im $\Sigma \propto \omega$ dependence dominates the whole momentum space. For high doping levels and high temperatures, Im $\Sigma \propto \omega^2$ shows up, starting from the antinodal region. By comparing with the hole-doped cuprates $La_{2-x}Sr_xCuO_4$ and $Bi_2Sr_2CaCu_2O_8$, we find a dichotomy of the scattering rate exists along the nodal and antinodal direction, which is ubiquitous in the cuprate family. Our work provides new insight into the strange metal state in cuprates.

The strange metal behavior is one of the most mysterious and important unsolved problems in high-temperature (high-$T_c$) superconductors[1-3]. These strange normal state properties are believed to be closely related to its high-temperature superconductivity[4-7]. Hence, uncovering the nature of strange metal is an essential step towards the final understanding of the mechanism of high-$T_c$ superconductivity. The most prominent characteristic of the strange metal is the linear-in-$T$ resistivity in contrast to the quadratic-$T$ resistivity in Fermi liquids. The linear-$T$ resistivity was reported up to several hundred kelvins and extends to low temperatures. It persists down to 1.5 K in the hole-doped cuprate La$_{2-x}$Sr$_x$CuO$_4$ (LSCO), to 40 mK in electron-doped cuprate Pr$_{2-x}$Ce$_x$CuO$_4$ (PCCO), and dominants the normal state in many other cuprates[8-10]. Among them, the electron-doped cuprate La$_{2-x}$Ce$_x$CuO$_4$ (LCCO) has very prominent strange metal behavior, with linear-in-$T$ resistivity persisting down to 20 mK, and a universal scaling law has been discovered recently with combinatorial films encompassing the entire overdoped range[2,4,5].

To unveil this anomalous property, the elementary information is the low-energy excitations behavior in energy-momentum space. For Fermi liquids, the quasiparticle lifetime is determined by the imaginary part of the self-energy Im $\Sigma(k, \omega)$ with $\omega^2$ dependence[11]. This quadratic energy dependence further guarantees the low-temperature quadratic-$T$ resistivity in conventional metals. Phenomenologically, the Im $\Sigma(k, \omega)$ of strange metal is supposed to have a linear energy dependence, also known as the marginal Fermi liquid (MFL) behavior[12-14]. In order to extract the momentum-dependent excitation information, we carried out a systematic study on the LCCO thin films using angle-resolved photoemission spectroscopy (ARPES) from the very underdoped regime to the highly overdoped regime. Previously, due to difficulties of growing the T′ phase single crystalline films and obtaining ultra clean and atomically flat surfaces, a full momentum-space ARPES study in LCCO is quite challenging[15-17]. In this work, by finely tuning the ozone/vacuum annealing procedure, we successfully perform ARPES experiments on the high-quality PLD-grown LCCO thin films[18], and provide a reliable way to extract the self-energy $\Sigma(k, \omega)$ and information of the single-particle scattering through Im $\Sigma(k, \omega)$[19].

The main results are summarized in the schematic phase diagram Fig. 1[5,20,21]. For LCCO, in the low doping regime at low temperatures, Im $\Sigma$ shows a homogenous behavior over the momentum space with Im $\Sigma \propto \omega$ (red lines in the bottom of the left panel). In the high doping regime at high temperatures, there is a crossover from Im $\Sigma \propto \omega$ to Im $\Sigma \propto \omega^2$ (blue lines), primarily showing up in the antinodal region. This dichotomy has also been observed in the hole-doped cuprates. For example, the overdoped LSCO shows the Im $\Sigma \propto \omega^2$ behavior in the nodal region while Im $\Sigma \propto \omega$ in the antinodal region, as plotted in the bottom of the right panel in Fig.1.

To support our observations, we first show the doping dependence of the momentum distribution curve (MDC) width (linewidth). The self-energy is related to MDC by Im $\Sigma(k, \omega)$ = $v_k\Gamma_k$, where $v_k$ is the velocity and $\Gamma_k$ is the half-width at half maximum (HWHM) of the MDC[22,23]. Using the refined ozone/vacuum annealing method, we obtain LCCO samples with the electron doping levels $n \sim 0.05$–0.23, from as-grown LCCO films with Ce concentration $x$ = 0.1, 0.15, and 0.19[18]. Since the antinodal region is plausibly affected by the kink effect and the stronger quasiparticle scattering rate, we mainly focus on the nodal direction[18,24-26]. Figure 2a shows the ARPES intensity plot of LCCO with the doping levels $n \sim 0.06$, 0.10, and 0.15. The cut in the momentum space is the same as cut 1 in Fig. 3b. The black line at each doping level indicates the band dispersion extracted from MDCs with a Lorentzian fit. Those black lines are straight, showing that no kink effect exists in the nodal region of LCCO at all doping levels, which is different from the antinodal region[26].

The doping dependence of linewidth as a function of $\omega$ at the nodal region is shown in Fig. 2b. The MDC widths at different doping levels are well-fitted by the $\omega$-linear dependence while clearly deviating from $\omega^2$. As a further demonstration, we use the power-law function $\Gamma(\omega) = a + b\omega^\alpha$ to fit the linewidths, and the power coefficient $\alpha$ goes to $\sim 1$ at various doping levels, as shown in Figs. 2c–2d [21,27]. It indicates the strange metal behavior in LCCO emerging from the nodal region, at least partially, if not all. Furthermore, we note that the slope of the linear fit is inversely correlated with the doping level in the overdoped region. This observation coincides with the scaling of the strange-metal scattering $T_c \sim (x_c-x)^{0.5} \sim A^{0.5}$ obtained by transport measurements, suggesting that the linear-in-$T$ resistivity may be an intrinsic property of LCCO at the nodal region[4]. Transport measurements contain contributions from interactions all over the momentum-space in the vicinity of $E_F$, while ARPES measures the single-particle scattering rate versus $\omega$ at a certain momentum[1,28,29]. For a fair comparison, it is necessary to measure the momentum dependence of the scattering rate by ARPES.

Figure 3 shows the momentum dependence of the MDC widths. Figure 3a is the ARPES spectra at different momentum spaces ($k_{//}$) along the cuts 1-6 indicated in Fig. 3b, the Fermi surface mapping at the doping level $n \sim 0.15$. Extracting the MDC widths by Lorentzian fitting, we plot them as a function of $\omega^2$ in Fig. 3c. The dashed power-law fitting lines are clearly deviated from linear, indicating that the exponential coefficient is less than 2, as presented in Fig. 3d. We note that from the nodal region to near the antinodal region, the linewidth is better described by the linear dependence $\Delta\Gamma \propto \omega$, where $\Delta\Gamma=\Gamma(\omega)-\Gamma(0)$. Only at the antinode, the coefficient jumps to around 1.5.

To further clarify the $\omega$-dependence of the single-particle self-energy, we performed temperature-dependent measurements with $n \sim 0.06$ (underdoped), 0.10 (optimally doped), and 0.15 (overdoped) samples. As seen in Figs. 4a–4b, in the underdoped and optimally doped

LCCO, the antinodal linewidth scales linearly with $\omega$ below 120 K and shows a linear to quadratic dependence at around 200 K. While the linear-in-$\omega$ non-Fermi liquid behavior persists to 200 K in the nodal region at $n \sim 0.10$. For $n \sim 0.15$ shown in Fig. 4c, the crossover from $\Delta\Gamma \propto \omega$ to $\Delta\Gamma \propto \omega^2$ occurs at an intermediate temperature range of 70–120 K, and the antinodal region has a more remarkable trend. It is intriguing that at high temperatures, the exponential coefficient α is not exactly at 2, but closer to 1.5, which is similar to the coefficient from transport measurements[5], where $R \propto T^{1.6}$. We summarize the doping, momentum, and temperature-dependent linewidths in Fig. 1. It can be simply viewed as that Im $\Sigma \propto \omega$ dependence dominates the whole momentum space at low doping levels and low temperatures. At high doping levels with high temperatures, Im $\Sigma \propto \omega^2$ shows up and is more prominent in the antinodal region.

The strange metal behavior is ubiquitous in the cuprate family, revealed from the comparisons of transport and spectroscopic data in the hole and electron-doped cuprates, we find that a dichotomy of the scattering rate exists along the nodal/antinodal direction. In $Bi_2Sr_2CaCu_2O_8$ (BSCCO), at the doping region where transport observes the linear-in-$T$ resistivity, the ARPES spectra show that the linear-in-$\omega$ mainly exists along the nodal direction[30-32]. It has been argued that the scattering rates for the in-plane transport are dominated by the behavior measured in the nodal region, owing to the pseudogap, the smaller Fermi velocity ($v_F$), and the larger scattering rate along the antinodal direction[12,28,33]. In the electron-doped LCCO, our spectroscopic experiments find that the nodal region is dominated by the linear-in-$\omega$ relation, which transforms to $\omega^{1.5}$ at higher temperatures in the overdoped regime. The antinodal single-particle self-energy has a faster transition from linear-in-$\omega$ to the Fermi-liquid-like $\omega^2$ dependence. The consistent results between BSCCO and LCCO suggest that the strange-metal behavior is mainly related to the nodal region. On the contrary, for the overdoped LSCO ($p$ ~0.28), Fermi liquid quasiparticles exist around the nodal region, whereas non-Fermi liquid excitations occur near the antinodal region[8,21]. Since LSCO and LCCO are the hole- and electron-doped counterparts of their parent material $La_2CuO_4$, it is astonishing to find a large discrepancy between them[34]. Understanding the dichotomy of the scattering rate in the nodal/antinodal region could be essential to unravel the mystery of the strange metal behavior and the high-$T_c$ superconductivity itself.

Phenomenologically, we can understand the disparity of the scattering rate in the hole- and electron-doped cuprates, based on the MFL theory, irrespective of the microscopic mechanism[12,13]. According to the MFL assumption, the single-particle self-energy is calculated to be:

$$\Sigma(\mathbf{k}, \omega) = \Sigma_1(\mathbf{k}, \omega) + i\Sigma_2(\mathbf{k}, \omega) = \lambda[\omega \log \frac{x}{\omega_c} - i\frac{\pi}{2}x], \qquad (1)$$

where $x \approx \max(|\omega|, T)$, and $\lambda$ is the coupling constant. By solving the Boltzmann equation, the conductivity is given by:

$$\sigma \propto \int_0^{\pi/2} N(\theta)v(\theta)M(\theta), \qquad (2)$$

$N(\theta)$ is the density of states (DOS) at the Fermi surface (FS) at angle $\theta$ and $v(\theta)$ is the Fermi velocity. $M(\theta)$ is related to the small-angle scattering, and the primary term is $v(\theta)$ by expanding it. Therefore, the k-dependent conductivity is proportional to $N(\theta)v^2(\theta)$.

The relationship is further affirmed by the Chambers' solution to the Boltzmann transport equation[35,36]:

$$\sigma = \frac{e^2}{4\pi^3} \oint d^2\mathbf{k} N(\mathbf{k}) v[\mathbf{k}(t=0)] \int_{-\infty}^{0} v[\mathbf{k}(t)] e^{t/\tau} dt, \qquad (3)$$

where the integral of $v$ is weighted by the probability of quasiparticles with a lifetime $\tau$ scattered after time $t$, and $N(\mathbf{k}) = 1/|\frac{\partial \varepsilon}{\partial \mathbf{k}}|$ is the DOS. The FS can be fitted by the effective tight-binding model:

$$\varepsilon = \mu - 2t[\cos(k_x a) + \cos(k_y a)] - 4t'\cos(k_x a)\cos(k_y a), \qquad (4)$$

Then we could estimate the contribution to the conductivity at a specific angle $\theta$ by calculating $N(\theta)v^2(\theta)$. The results for overdoped LSCO ($p \sim 0.28$), BSCCO ($p \sim 0.24$), and LCCO ($n \sim 0.15$) at the nodal and antinodal regions are shown in Tab. 1. Since the overdoped LSCO sample has the extended van-Hove singularity and the FS segment is absent at the antinode $\theta = 0°$, thus we compare the conductivity near the angles of $\theta \sim 15°$ and $45°$[21,37].

From Tab. 1, we note that the ratio of the conductivity at the antinode to the node region $\sigma(\text{an})/\sigma(\text{node})$ is the largest in LSCO, suggesting that the antinodal region of overdoped LSCO has a considerable contribution to conductivity among other cuprates. It partially explains the striking feature that the linear-$\omega$ dependence principally comes from the antinode. From the table, we assume that it is mainly due to the large density of states exceeding the van-Hove singularity at the antinodal region[37,38]. The ratio is modest and comparable in LCCO, where the node has a slightly larger contribution to the conductivity than the antinode, which is consistent with our ARPES experiments. In our observations, the nodal and antinodal region crossover from $\omega$ to $\omega^2$ dependence and the nodal region is primarily dominated by the strange linear behavior. While in BSCCO, the magnitude of $\sigma(\text{an})/\sigma(\text{node})$ is smaller than the values in LSCO and LCCO, mainly influenced by the minor Fermi velocity at the antinode, suggesting the linear-$T$ transport scattering rate corresponding to the linear-$\omega$ single particle scattering rate at the nodal region in BSCCO. Due to the large gap opening, $v(\text{an})$ is estimated around the minimum gap energy[39]. It is worth noting that at $\theta \sim 0°$ (antinode), the Fermi velocity nearly vanishes, and the density of states would be gapped, contributing much less to the conductivity in BSCCO[28,39,40]. From this respect, we could qualitatively relate the linear-$T$ strange metal phenomenon to the Fermi liquid breakdown in the momentum space, which could enlighten the understanding of the dichotomy of the scattering rate observed in both hole- and electron-doped cuprates.

Microscopically, numerous theoretical scenarios have been proposed to explain the strange metal behavior, especially the Planckian dissipation, quantum criticality associated with AF spin fluctuations, etc[1,3,29,32,41]. How these theories relate to ARPES observations is still an open question. For Planckian dissipation theory, the scattering rate is limited by the Planckian limit as $\hbar/\tau = \alpha k_B T$, $\hbar$, $\tau$, and $k_B$ are the reduced Planck constant, relaxation time, and Boltzmann constant, $\alpha$ is of order unity[29,42]. But the momentum information of this theory is lacking. Previous work also shows the inverse of the relaxation time $1/\tau$ for LCCO drastically exceeds the putative Planckian limit, once the $T^2$-resistivity establishes[2,4]. On the other hand, in the marginal Fermi liquid phenomenology associated with the ($T$, $\omega$)-linear self-energies, the common interpretation is centered on quantum criticality. It is a prevailing consensus that the strange metal in the cuprates is tied to the AF spin fluctuations correlated with the quantum critical point or an extended quantum criticality[2,41,43,44]. Following this scenario, we try to link the AF fluctuation and the dichotomy of the scattering rate. The black dashed line in the last column of Tab. 1 is the antiferromagnetic zone boundary (AFZB). As in LCCO, the nodal regions remaining linear-in-$\omega$ are connected by the ($\pi$, $\pi$) AFZB, which is contradictory to BSCCO, where the AFZB is more closely related to the antinodal region. This situation is even more mysterious in overdoped LSCO with square-like FS, where the ($\pi$, $\pi$) AFZB could not distinguish the $\omega$ and $\omega^2$ segments. Hence, how to microscopically understand this dichotomy is still challenging, which calls for further theoretical and experimental investigations.

In summary, we perform doping, momentum, and temperature-dependent ARPES measurements and reveal the evolution of the strange metal scattering in the momentum space of electron-doped cuprate LCCO. By comparing with hole-doped cuprates LSCO and BSCCO, we discover the dichotomy of the scattering rate in the nodal and antinodal regions. The marginal Fermi liquid theory may offer a qualitative interpretation considering the DOS and Fermi velocity, but a complete microscopic understanding is still missing. Given the lack of a full momentum-space description for the Fermi liquid breakdown, we hope our work will stimulate more theoretical and experimental work to unveil the strange metal phenomenon and high-$T_c$ superconductivity itself.

## Method

High-quality LCCO films with Ce concentration $x \sim 0.1, 0.15$ and $0.19$ are grown by pulsed laser deposition. By the elaborate two-step ozone/vacuum annealing method, we can obtain flat and clean surfaces for ARPES measurements, with electron doping level $n \sim 0.05-0.23$. For $n \sim 0.05-0.19$, ARPES measurement were performed with a Scienta R4000 analyzer and a VUV Helium light source. We use He-II$\alpha$ (40.8 eV) resonant light, and the base pressure of the ARPES system is $\sim 4 \times 10^{-11}$ torr. All measurements were carried out at 30 K unless specially stated. For $n \sim 0.23$, ARPES spectrum was measured at $\sim 20$ K at the Dreamline beamline of the Shanghai Synchrotron Radiation Facility (SSRF) with a Scienta Omicron DA30L, photon energy with 55 eV was used.

## Acknowledgements

We thank J.X.Z. and Z.Y.W. for valuable discussions. This work was supported by the grants from the Natural Science Foundation of China (12141402), the Ministry of Science and Technology of China (2017YFA0403401), the Chinese Academy of Sciences (XDB25000000).

## Author contributions

C.Y.T. and Y.J.S. designed the experiments. Z.F.L. and K.J. grew the thin films, C.Y.T. performed ARPES measurements with the assistance of X.C.G., S.Y.G., Z.C.R., J.Z., Y.B.H., and T.Q. C.Y.T. analyzed the ARPES data with help from Y.G.Z. and J.Y.G. K.J. and X.L.F. provided the theoretical advices. C.Y.T. and H.D. wrote the manuscripts with help from all the other authors. K.J., K.J., Y.J.S., and H.D. supervised the project.

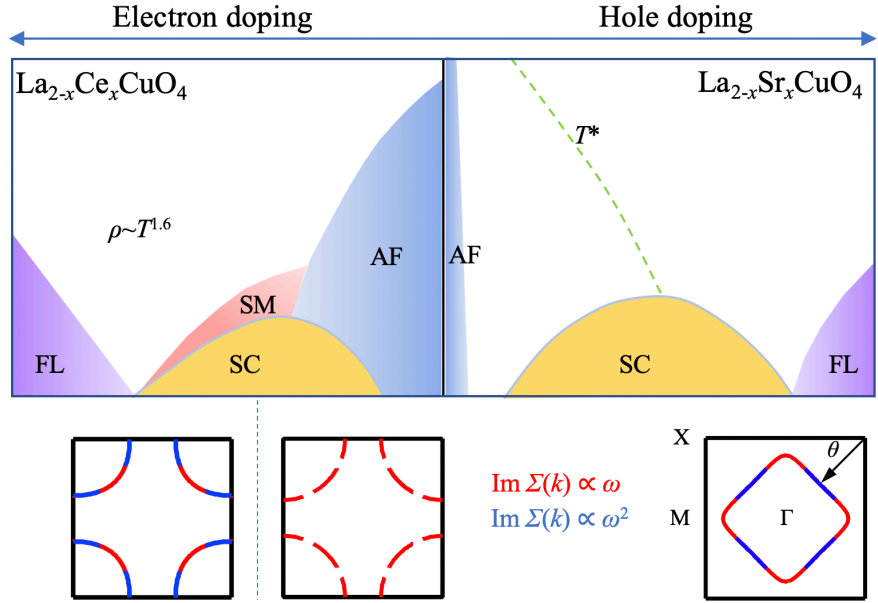

**Fig. 1 | Schematic phase diagram of the hole- and electron-doped cuprate La$_{2-x}$Sr$_x$CuO$_4$ (LSCO) and La$_{2-x}$Ce$_x$CuO$_4$ (LCCO).** AF, antiferromagnetism; SC, superconductivity; FL, Fermi liquid; SM, strange metal. In LSCO with the doping level $p = 0.28$, the non-Fermi liquid state appears through an angular breakdown. The arcs of Fermi liquid quasiparticles Im $\Sigma \propto \omega^2$ (blue lines in the bottom of the right panel) separates by gapless non-Fermi liquid excitations with Im $\Sigma \propto \omega$ (red lines)[21]. While in the electron-doped copper oxide La$_{2-x}$Ce$_x$CuO$_4$ (LCCO), the quasiparticle self-energy crossover from Im $\Sigma \propto \omega$ to Im $\Sigma \propto \omega^2$ at higher electron doping levels and higher temperatures, starting from the antinode (bottom of the left panel). The Fermi arcs in the underdoped region of electron doping represent the existence of hotspot[18]. The phase diagram of LCCO and LSCO are adapted from refs. 6 and 21, respectively.

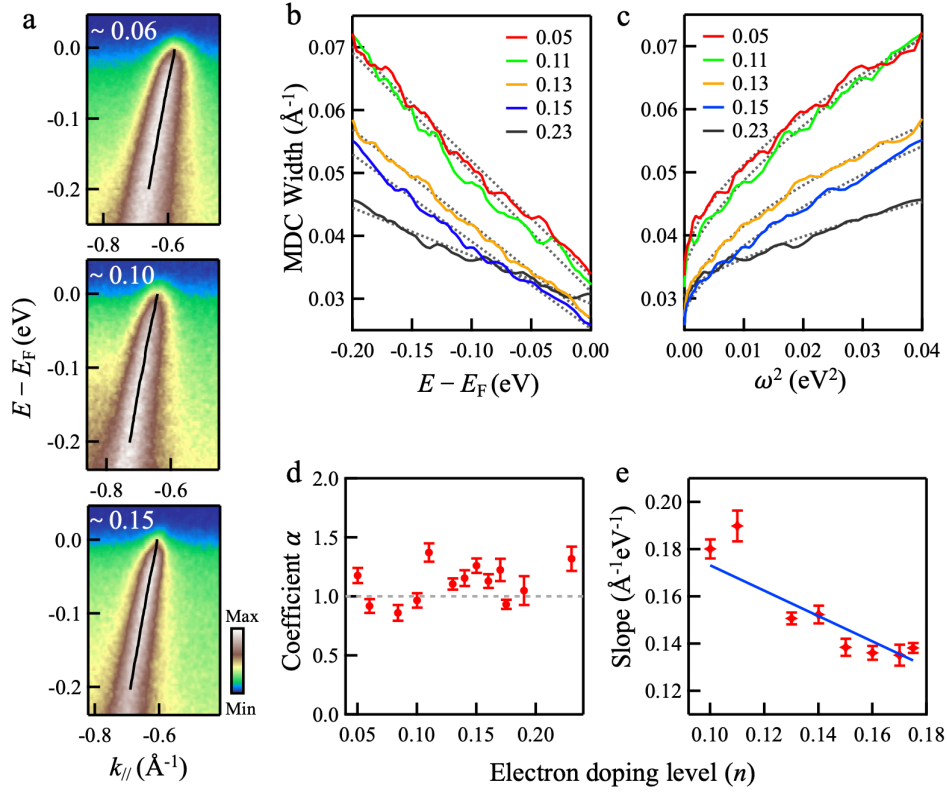

**Fig. 2 | Doping dependence of nodal dispersion of LCCO. a,** The nodal band dispersion of LCCO with electron doping levels $n \sim 0.06$, 0.10, and 0.15. **b,** Half-width at half maximum of the momentum distribution curves (MDC width) versus excitation energy $\omega = E - E_F$ at the nodal region for various doping levels. The dashed lines are the linear fits to the linewidth. **c,** Nodal linewidth plotted versus $\omega^2$. Dashed lines are the power-law fitting with $\Gamma(\omega) = a + b\omega^\alpha$, the power coefficient $\alpha$ is shown in **d. e,** The slope of the linear fit of the linewidth as a function of $n$ in the overdoped region. The error bar is twice the stand deviation from the fitting procedure. The nodal band dispersions at all doping levels are measured at 30 K.

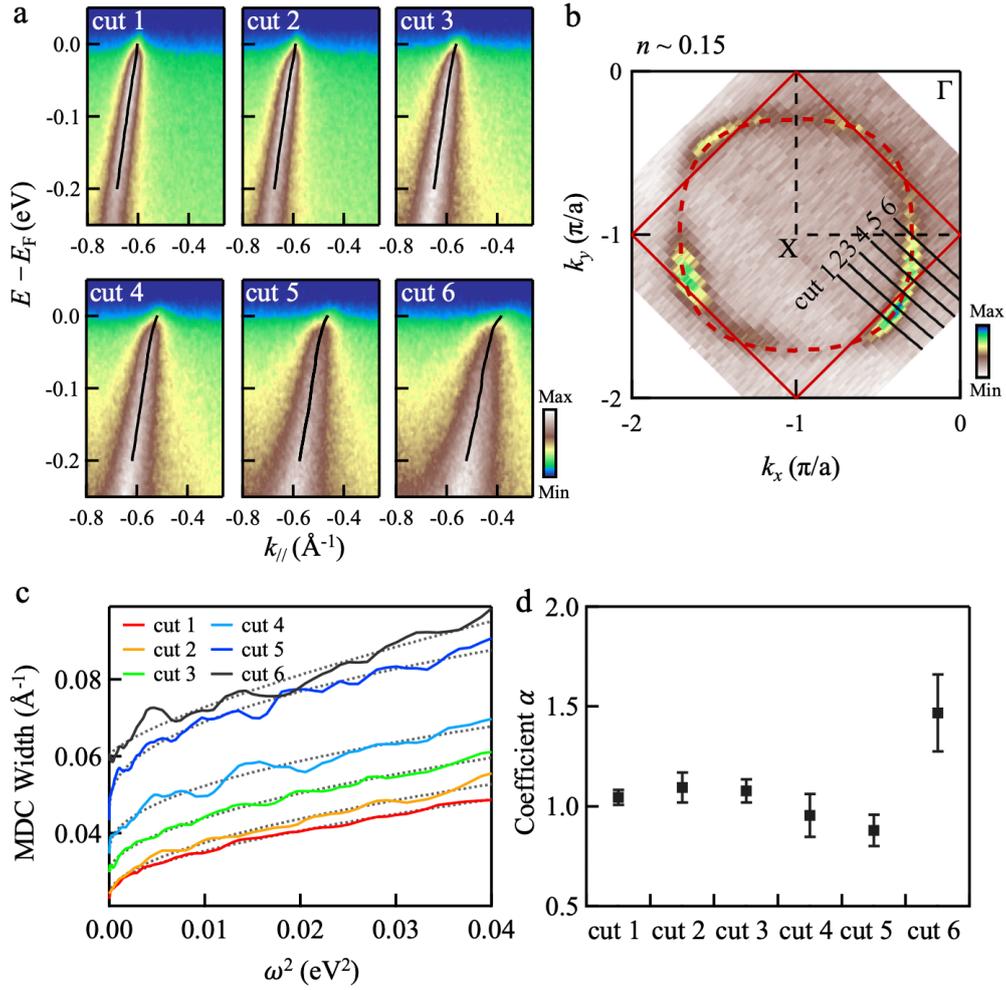

**Fig. 3 | Momentum dependence of the MDC widths. a,** Band dispersions of cuts 1-6, as indicated in **b**. The black lines are the peaks of the MDCs extracted by Lorentzian fitting. **b,** Fermi surface (FS) map fitted by the tight-binding model, the doping level is estimated to be $n \sim 0.15$. Band dispersions and FS mapping are measured at 10 K and 30 K, respectively. **c,** MDC width versus $\omega^2$, using power-law function to fit the data (dotted lines), and the exponential coefficient is plotted in **d**.

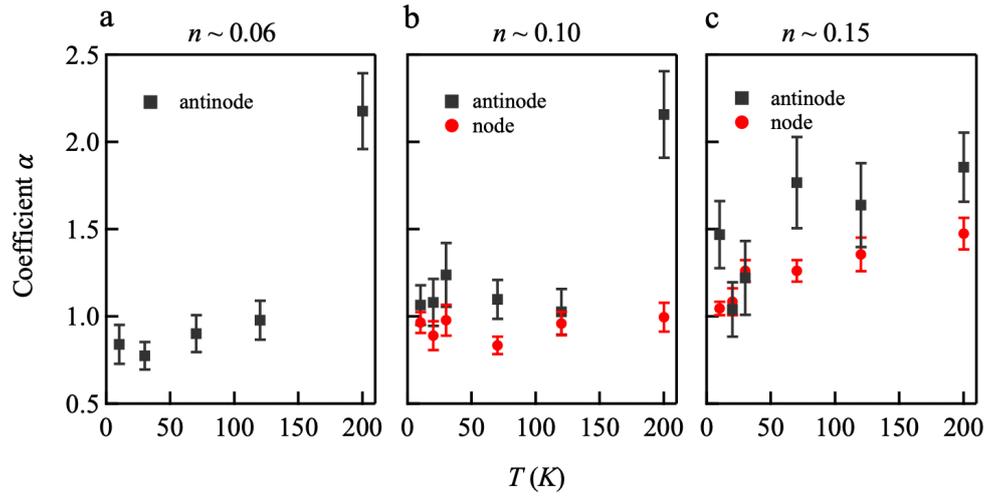

**Fig. 4 | Temperature-dependence of the MDC widths. a-c,** The exponential coefficient α of the power-law fitting function $\Gamma(\omega) = a + b\omega^\alpha$ for $n \sim 0.06$, 0.10, and 0.15, respectively, at different measured temperatures.

**Tab. 1 | Dichotomy of the scattering rate in hole- and electron-doped cuprates.** The Fermi velocity $v(\theta)$ is corrected by the measured angle. Antinodal (an) and nodal is the region near the angle $\theta \sim 15°$ and $45°$, respectively. The density of states $N(\theta)$ is calculated as $1/|\frac{\partial \varepsilon}{\partial \mathbf{k}}|$. And therefore, the contribution to the conductivity is proportional to $v^2(\theta)N(\theta)$. The FS topology is plotted in the last column, where red and blue FS sections represent linear-$\omega$ and $\omega^2$ dependence, and the black dashed line indicates the antiferromagnetic zone boundary. The data of LSCO is adapted from ref. 21, the data of BSCCO is adapted from refs. 39 and 40.

| | $\dfrac{v(\text{an})}{v(\text{node})}$ | $\dfrac{N(\text{an})}{N(\text{node})}$ | $\dfrac{\sigma(\text{an})}{\sigma(\text{node})}$ | FS topology |
|---|---|---|---|---|
| LSCO ($p \sim 0.28$) | 0.53 | 2.16 | 0.607 | 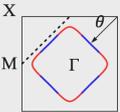 |
| BSCCO ($p \sim 0.24$) | 0.28 | 1.21 | 0.095 | 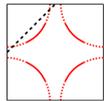 |
| LCCO ($n \sim 0.15$) | 0.61 | 0.78 | 0.290 | 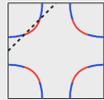 |